\ifdefined\pdfoutput\pdfoutput=1\fi
\documentclass[11pt,a4paper]{article}

\usepackage[T1]{fontenc}
\usepackage[utf8]{inputenc}
\usepackage{amsmath}
\usepackage{newtxtext}
\usepackage{newtxmath}
\DeclareUnicodeCharacter{00B5}{\ensuremath{\upmu}}
\usepackage[a4paper,margin=25mm]{geometry}
\usepackage{graphicx}
\usepackage[font=small,labelfont=bf,labelsep=period,justification=justified,singlelinecheck=false,skip=6pt]{caption}
\usepackage[nosort]{cite}
\usepackage{placeins}
\usepackage{microtype}
\usepackage{xurl}
\usepackage[hidelinks,bookmarksopen=true]{hyperref}
\hypersetup{pdftitle={Allight: A Perfect Photon Collector},pdfauthor={Dan Porat, Ran Finkelstein, Tal Carmon}}

\makeatletter
\renewcommand{\@biblabel}[1]{#1.}
\makeatother
\renewcommand{\refname}{References and Notes}


\begin{document}

\begin{center}
  {\LARGE\bfseries Allight: A Perfect Photon Collector\par}
  \vspace{1.2em}
  {\large Dan Porat\textsuperscript{1,2}, Ran Finkelstein\textsuperscript{2*}, Tal Carmon\textsuperscript{1*}\par}
  \vspace{0.9em}
  {\small
  \textsuperscript{1}Electrical and Computer Engineering, Tel Aviv University; Tel Aviv 69978, Israel.\par
  \textsuperscript{2}School of Physics and Astronomy, Tel Aviv University; Tel Aviv 69978, Israel.\par
  \vspace{0.6em}
  \textsuperscript{*}Corresponding authors. Email:
  \href{mailto:ranf@tauex.tau.ac.il}{ranf@tauex.tau.ac.il} (R.F.);
  \href{mailto:total@tauex.tau.ac.il}{total@tauex.tau.ac.il} (T.C.)\par}
\end{center}

\begin{abstract}
\noindent Collecting photons from quantum emitters remains a key limitation in
light-matter interaction because conventional lenses collect light only
from one direction. Here, we introduce a perfect light collector by
inverse-designing a surrounding dielectric structure that encloses the
emitter to increase photon-collection efficiency toward unity.
Decomposing the device into several refractive interfaces enables
emission collection from the full 4$\pi$ solid angle with adiabatic
reshaping into a collimated beam. We experimentally demonstrate, in two
dimensions, the ray bending that collimates omnidirectional emission.
Our collector transforms current paradigms for highly efficient photon
collection from narrowband, cavity-based techniques into experimentally
robust, broadband, non-resonant dielectric devices.
\end{abstract}

\makeatletter\def\toclevel@subsection{1}\makeatother
\bigskip

Collecting the light radiated by a localized emitter and directing it
into a collimated beam is a fundamental task in optics. Conventional
collection optics intercept only a fraction of the emitted radiation,
leaving light outside their angular acceptance (numerical aperture)
unused. Improving collection would benefit a wide range of applications,
ranging from fluorescence detection to single-photon sources
\cite{Claudon2010,Maiwald2012,Somaschi2016,Huang2019,Carter2024}. It is particularly important for photonic
interconnects, which use light to distribute entanglement between
spatially separated matter-based qubits \cite{Kimble2008,Reiserer2015,Wehner2018}. Such
interconnects provide a route to scaling quantum computers beyond
individual quantum processing units and thereby enable scaling qubit
count into the regime required for useful fault-tolerant computing.
Recent demonstrations include quantum logic between distant network
modules \cite{Daiss2021}, distributed quantum computing across an optical link
\cite{Main2025}, photon-interfaced multiqubit registers \cite{Canteri2025}, remote
entanglement between solid-state quantum memories \cite{LagoRivera2021,Knaut2024}, and entanglement between single atoms over telecom fiber
\cite{VanLeent2022}. Beyond computation, distributing entanglement enables
applications in quantum cryptography \cite{Wengerowsky2018,Yin2020} and
distributed sensing \cite{Guo2020,Malia2022,Bate2025}. All these applications
require optical interfaces that collect photons efficiently and deliver
them in modes suitable for transmission and interference.

As optical fibers and single-photon detectors approach unit efficiency,
photon collection becomes an increasingly prominent bottleneck in
quantum interconnects. Specifically, single-photon detectors can now
achieve system efficiencies reaching unity at telecom wavelengths
\cite{Reddy2020,Chang2021}, and transmission losses are small over short
optical links. By comparison, collecting emission from a localized
matter-based qubit into a useful optical mode remains extremely
inefficient. For example, reported photon-delivery probabilities reach
only the percent level for trapped ions and neutral atoms \cite{OReilly2024,Safari2026}. In this regime, collection losses dominate the optical loss
budget, suppressing remote-entanglement generation rates \cite{Stephenson2020,Krutyanskiy2023} and constraining the size and operating speed of distributed
quantum systems \cite{Aqua2026}. Such losses compound when entanglement
generation requires detecting photons from both nodes. In two-photon
heralding protocols, the success probability scales as
$\eta^{2}$ \cite{Raskop2026}, where $\eta$ is the
photon-collection efficiency at each of two identical interfaces. Even a
lens with numerical aperture 0.8, as impressively demonstrated in a high
numerical aperture (NA) trapped-ion interface \cite{Carter2024}, subtends only
20\% of the full solid angle. For an isotropic emitter, increasing this
collection fraction toward unity would enhance the two-photon heralding
rate by up to a factor of 25, with other efficiencies and the attempt
rate held fixed. Improving collection would reduce the waiting time for
entanglement between remote processors, with exponentially larger
benefits in protocols that require collecting several photons within the
same attempt.

Several approaches address this optical-interface challenge. A prominent
approach integrates an optical cavity around a quantum emitter or qubit.
Optical cavities enhance coupling to selected modes and have been
integrated with trapped ions and neutral atoms \cite{Steiner2013,Kato2015,Will2021,Canteri2025}, while cavity-enhanced quantum-dot sources combine high
brightness with coherent single-photon emission \cite{Tomm2021}. However,
such resonant interfaces depend on spectral and spatial mode matching,
require active stabilization \cite{Steiner2013,Kato2015,Carmon2005}, and may
induce stray fields that hamper qubit coherence. In contrast,
conventional lenses offer robust, broadband, and technically simple
operation, but their light-collection efficiency is limited by their NA.
As an example, even a state-of-the-art dry objective with NA of 0.95
would collect and collimate only 34\% of the light emitted by a point
source. Attempts to realize efficient collectors for quantum emitters
include high-numerical-aperture lenses \cite{Carter2024}, immersion meta-lenses
\cite{Huang2019}, and deep parabolic mirrors \cite{Maiwald2012}. Yet the collection
efficiency of such devices is still far smaller than unity, and their
integration with real quantum processors remains an outstanding
challenge.

A broader perspective comes from metamaterials \cite{Bomzon2002,Yu2011,Zheludev2012},
transformation optics \cite{Pendry2006,Leonhardt2006}, and freeform optics
\cite{Bauer2018,Rolland2021}, which have shown how spatially structured
optical media can prescribe light propagation in unconventional
geometries. Such approaches include the optical black hole proposed by
Narimanov and Kildishev, which uses a spatially varying dielectric
response to direct incident light toward an absorbing core over a broad
bandwidth and from all directions \cite{Narimanov2009}. These ideas motivate our
complementary objective for quantum interfaces: surrounding an emitter
with an optical structure that redirects its omnidirectional radiation
into a collimated output beam.

\begin{figure}[b]
  \centering
  \includegraphics[width=0.82\linewidth]{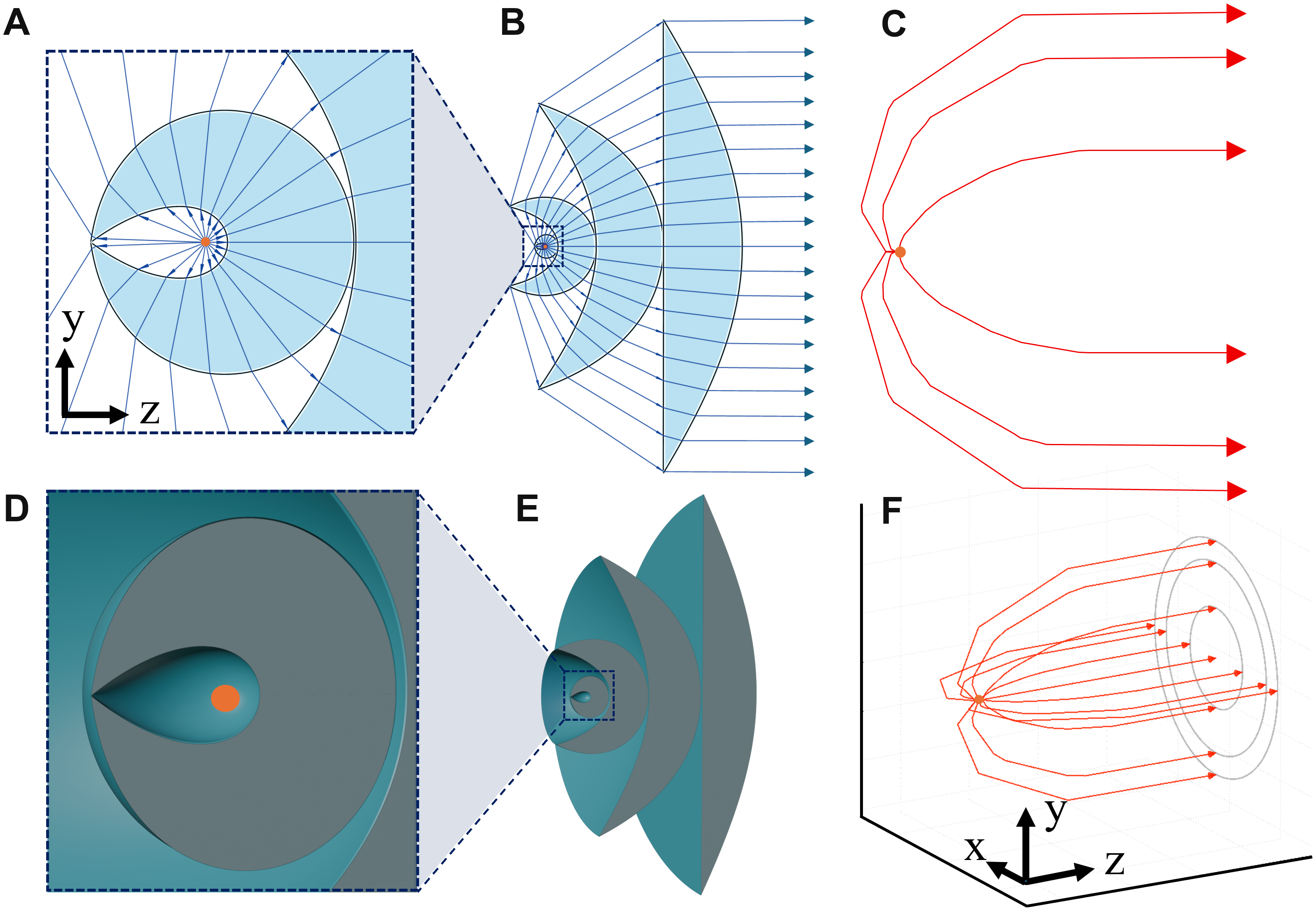}
  \caption{\textbf{Collector design.} \textbf{(A)} The first element of our
  collector is a dielectric structure fully surrounding an intended
  quantum emitter (orange). This first surrounding element is the key
  element in this work. \textbf{(B)} Additional nested elements are
  designed to fully collimate the light. Note that even the ray initially
  propagating to the left makes a U-turn and aligns along the $z$-axis (the
  optical axis). \textbf{(C)} Ray tracing showing refraction at the
  air-glass interfaces of each element. \textbf{(D} to \textbf{F)} Same as
  (A) to (C), in three dimensions. The cut (gray) in (D) and (E) is only
  for visualization purposes and makes the inner parts visible. Refractive
  index $n = 2.003$.}
  \label{fig:design}
\end{figure}

Here, we introduce a surrounding-dielectric architecture designed to
collect emission over the full 4$\pi$ solid angle. We call our
collector Allight, from ``all'' and ``light'', because it gathers nearly all
the light an emitter radiates in all directions into one collimated
beam. An inner dielectric element encloses the intended emitter
position, while successive refractive interfaces progressively turn rays
toward a common collimated output direction (Fig.~\ref{fig:design}).
We inversely design the
interface profiles to map light from a point emitter, such as an ion or
an atom, into a collimated beam that fully reaches an optical detector.
We do so by distributing the required deflection across multiple
dielectric interfaces. Our resulting geometry implements full-angle
light redirection using discrete interfaces between homogeneous
dielectric regions, offering a realistic route toward near-unity
collection with broadband, non-resonant operation. We develop the
rotationally symmetric three-dimensional geometry and test its
ray-steering principle in a fabricated two-dimensional prototype. Using
laser beams at different launch positions, we visualize trajectories
through the structure and compare them with the designed paths. These
measurements demonstrate the large-angle redirection underlying the
collector and provide an experimental basis for developing surrounding
optical interfaces for quantum emitters. Finally, we discuss how such a
collector can be integrated with various quantum platforms based on
light-emitting material qubits. We show how the common dipolar emission
pattern can be utilized to introduce large free-space access into the
internal volume of the surrounding collector, enabling efficient qubit
shuttling while retaining near-unity collection efficiency.

\subsection{Collector inverse design}
\label{sec:design}

Our main design criterion is that all rays emitted from a point-like
source, in any direction, emerge from the collector parallel to each
other, as one collimated beam. To collimate light emitted in all
directions, we start with a dielectric element that fully surrounds the
emitter (Fig.~\ref{fig:design}A, orange). This element and all other elements (Fig.~\ref{fig:design}B)
are axisymmetric around the optical axis (the $z$-axis in Fig.~\ref{fig:design}A).
Because of this axial symmetry, our two-dimensional drawing (Fig.~\ref{fig:design}, A
to C) holds for every plane that contains the axis. Hence, our
surrounding dielectric structure also collects and collimates rays in
each such plane (Fig.~\ref{fig:design}, D to F). We perform constrained optimization,
following the criterion that all rays should be tilted gradually, or
adiabatically, along their path. This gradual tilting would improve
transmission by lowering reflection at each interface and making
antireflective coating easier, as well as relaxing the tolerance to
fabrication and alignment errors. Indeed, our calculation shows that
with currently available fabrication and alignment tolerances ($\lambda/2$
aspheric surfaces; $\lambda/4$ alignment), the collection efficiency drops by
less than 0.1\% up to 3~m from the collector (fig.~\ref{fig:S3}). With
antireflection coatings that transmit 99.99\% per interface, the eight
interfaces together lose less than 0.1\% (see \hyperref[sec:fabricability]{Materials and Methods,
``Fabricability of a three-dimensional Allight collector''}). In the
example of Fig.~\ref{fig:design}, eight dielectric interfaces (composed of four
dielectric elements) tilt the rays. Accordingly, our optimization target
is for each interface to tilt the ray by one-eighth of its total
required tilt. For example, for a ray emitted directly away from the
detector, this target divides the 180° U-turn into eight successive
tilts of 22.5° each. With these design criteria, we inversely design the
geometry: for a fan of rays that continuously cover all emission
directions (the full 4$\pi$ solid angle), we compute the geometry of the
interface required to collimate each ray along the optical axis. Our
physical model makes no approximation, as it relies only on
Snell's law of refraction. We wrote the Allight design
software for this study. Its source code and compiled version are
available on GitHub (see \hyperref[sec:inverse-design]{Materials and Methods,
``Inverse-design algorithm''}), and figs.~\ref{fig:S1}
and~\ref{fig:S2} describe its algorithm. The output of this software is a CAD
(computer-aided design) file of the elements' geometry.
Figure~\ref{fig:design} contains such a design for four elements, each having a
refractive index $n = 2.003$. In the next section, we design and
fabricate a two-dimensional version of this design as a prototype for experimental
characterization.

\subsection{Collector fabrication and experimental ray tracing}
\label{sec:experiment}

\begin{figure}[tbp]
  \centering
  \includegraphics[width=0.72\linewidth]{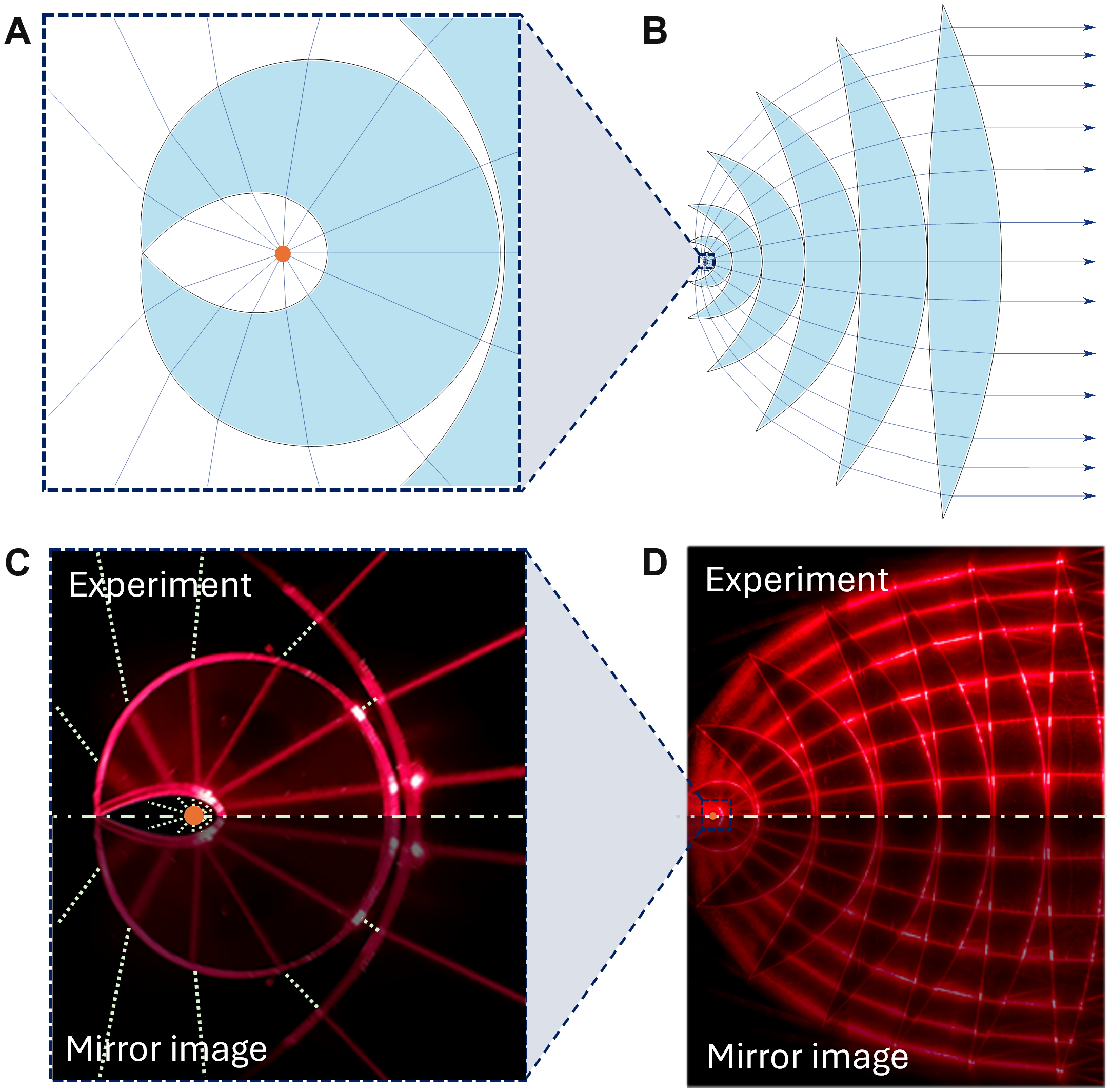}
  \caption{\textbf{Two-dimensional experimental proof of concept. (A} and
  \textbf{B)} Collector design for polycarbonate ($n = 1.57$); (\textbf{A})
  magnifies the region marked by the blue dashed box in (B). \textbf{(C}
  and \textbf{D)} The fabricated device, following the same design, with
  experimental ray tracing observed through scattering of laser beams ($\lambda =
  650$~nm) launched at different positions parallel to the optical axis;
  all beams converge to the focus (orange dot). (C) repeats the experiment
  of (D) with an enlarged replica of the region marked by the blue dashed
  box, to improve resolution. White dotted lines are a guide to the eye.
  In (C) and (D), only the upper half was fabricated; the lower half is a
  mirror image about the optical axis (dashed-dotted line), shown dimmed.}
  \label{fig:experiment}
\end{figure}

Invisibility cloaks were designed for three dimensions \cite{Pendry2006} but
first demonstrated in two \cite{Schurig2006,Smolyaninov2009}. We follow the same
route: here we design a two-dimensional collector (Fig.~\ref{fig:experiment}, A and B) from
polycarbonate, a transparent polymer with refractive index $n = 1.57$. We
fabricate the parts using a computer numerically controlled (CNC)
milling machine, followed by polishing.

Our collector maps between a collimated beam and a spherical wave, such
as the one a point-like emitter radiates. We verify this mapping by
launching laser beams ($\lambda = 650$~nm) parallel to the optical axis at
different positions across the collimated aperture and photographing the
residual scattering that marks each ray path. Because the collector is
mirror-symmetric about the optical axis, its top half contains every
designed ray path; we therefore fabricate this half and show its mirror
image to display the full profile (Fig.~\ref{fig:experiment}, C and D). The observed
trajectories match the designed ones across the full angular range: each
beam follows its designed path through successive interfaces, including
the rays that turn back by 180°, and converges to the focal point.

Revolving this two-dimensional profile around the optical axis would yield the
three-dimensional collector, which would be able to reach near-unity
collection efficiency of light from a quantum emitter.

\subsection{A route for integration with qubit arrays}
\label{sec:integration}

\begin{figure}[tbp]
  \centering
  \includegraphics[width=0.62\linewidth]{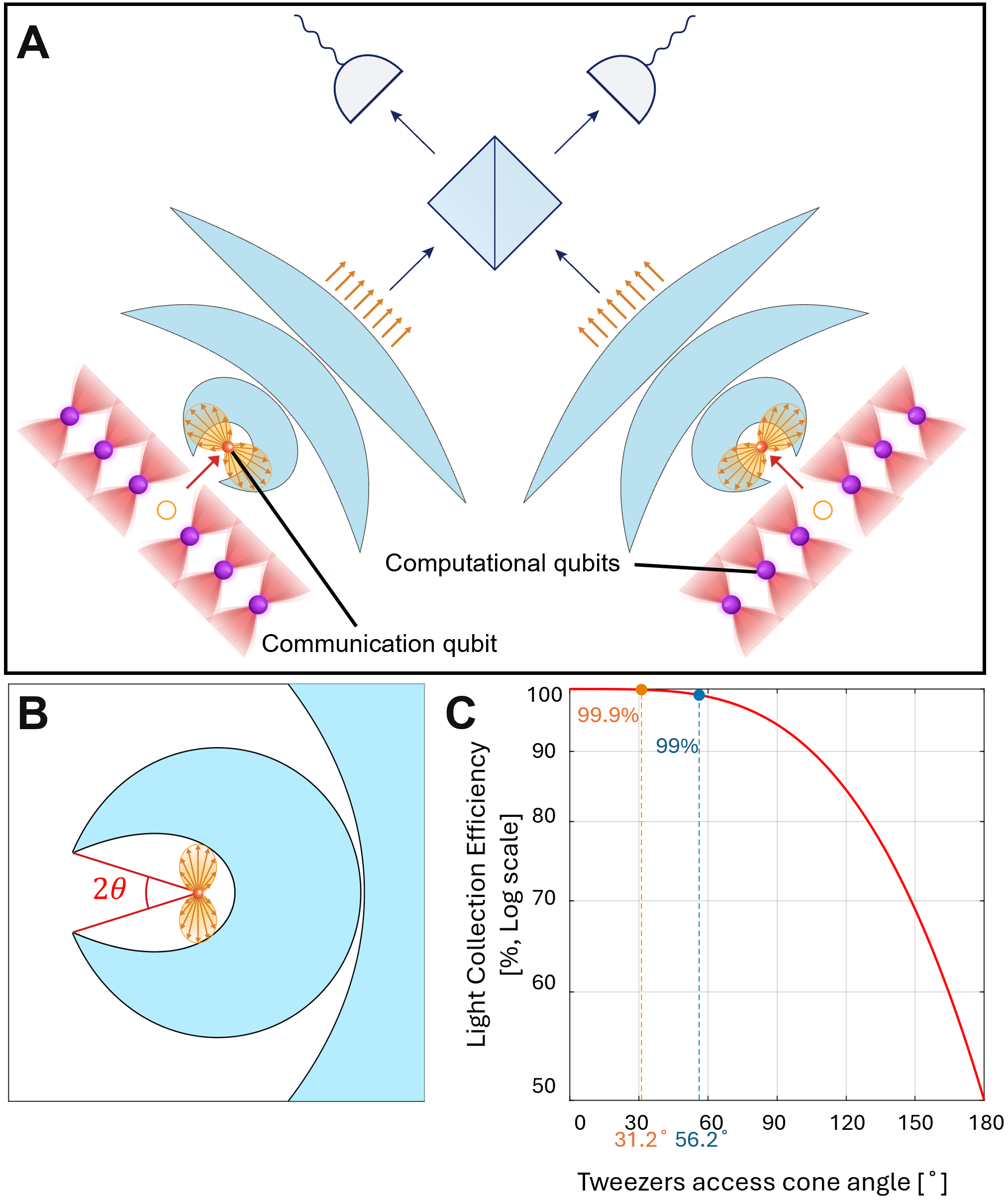}
  \caption{\textbf{Applying perfect collectors in a quantum network. (A)}
  Vision for a scalable quantum network where the remote-entanglement rate
  is significantly enhanced between communication qubits in remote nodes.
  In each node, local gates are performed between computational and
  communication qubits, where the latter can be shuttled into an internal
  void in a perfect collector where enhanced photon collection is used to
  herald remote entanglement with a communication qubit in a remote node.
  \textbf{(B)} A conical hole of apex angle $2\theta$ along the dipole
  axis, where the doughnut-shaped dipole emission (orange) is
  minimal, allows delivery of a communication qubit into the collector.
  \textbf{(C)} Collection efficiency $\eta$ (Materials and Methods, Eq.~\ref{eq:eta-theta})
  as a function of the cone apex angle $2\theta$, where $\theta$ is the half-angle of
  the cone and $\mathrm{NA} = \sin\theta$ its numerical aperture; $\eta$ remains at
  least 99\% for apex angles up to 56.2° ($\mathrm{NA} = 0.47$).}
  \label{fig:network}
\end{figure}

A fully enclosing collector leaves no free-space access to the emitter,
whereas trapping, cooling, and especially shuttling of qubits rely on
free-space entry. Such access enables shuttling (Fig.~\ref{fig:network}A) via, e.g.,
optical or electrical moving traps \cite{Barredo2016,Kaufman2021}. We introduce
this free-space path by exploiting the dipolar emission pattern common
to atomic and solid-state qubits, which has a doughnut-shaped profile
(Fig.~\ref{fig:network}B) with zero emission along the dipole-moment axis. Aligning the
dipole with the optical axis and leaving a conical opening of apex angle
$2\theta$ around it (Fig.~\ref{fig:network}B) sacrifices only the weak emission along the
dipole direction. For an opening of half-angle $\theta$, the collected fraction
is $\eta = (2 + 3\cos\theta - \cos^{3}\theta)/4$ (Materials and Methods, Eq.~\ref{eq:eta-theta}), so
the loss grows only as the fourth power of the opening's numerical
aperture, $\mathrm{NA} = \sin\theta$. The collection efficiency remains 99.9\% for a
31.2° apex opening and 99\% for a 56.2° opening (Fig.~\ref{fig:network}C); the latter
admits optical tweezers with a numerical aperture of about 0.47, on the
order of that needed for state-of-the-art mobile qubit processors.
Through this opening, a communication qubit can be delivered to the
collector focus following local entangling operations with computational
qubits held in a neighboring qubit array \cite{Bluvstein2022,Finkelstein2024}. Photons
emitted by the communication qubits of two such modules are collimated,
interfered on a beam splitter and detected, heralding entanglement
between the modules \cite{Barrett2005,Moehring2007}. Because each module
contributes one collected photon per attempt, the near-unity collection
enters the heralding rate squared, and it compounds further as more
modules are linked.

In conclusion, we have introduced a surrounding-dielectric architecture
designed to redirect emission over the full solid angle into a
collimated beam and experimentally demonstrated its ray-steering
principle in a two-dimensional prototype. By distributing the required
deflection across inversely designed refractive interfaces, this
approach offers a route toward near-unity photon collection without
resonant enhancement or active resonance tuning. Efficient collection
could benefit applications ranging from fluorescence detection to
quantum light sources and photonic interconnects. For quantum networks,
approaching unity would remove collection loss as a major penalty in
remote-entanglement generation, increasing entanglement distribution
rates and reducing the time that quantum memories must preserve
coherence while awaiting successful connections. These gains would
support the assembly of larger distributed entangled states and faster
operations between quantum processors, enabling the large system sizes
required for utility-scale fault-tolerant quantum computing \cite{Reiher2017,Gidney2021}. Our calculations and two-dimensional proof of concept,
together with mature three-dimensional fabrication techniques
\cite{Bauer2018,Rolland2021}, pave the way for near-term realization of
the full collector. The proposed integration geometry provides access
for introducing and controlling an emitter, benefiting coupling to
trapped quantum emitters and efficient quantum interconnects.

\FloatBarrier
\makeatletter\def\toclevel@subsection{2}\makeatother
\section{Materials and Methods}
\label{sec:methods}

\subsection{Inverse-design algorithm}
\label{sec:inverse-design}

We calculate the collector geometry with an inverse-design algorithm
implemented in a custom MATLAB routine. We assume an emitter with an
axially symmetric radiation pattern, such as an isotropic emitter or a
dipole aligned with the optical axis. We sample the emission with a fan
of rays distributed uniformly over the emitter's solid angle, with polar
launch angles $\alpha_{0}$ ranging from 0° (along the optical axis) to 180°
(against it). We require every ray to exit parallel to the optical axis
after crossing $L = 2N$ refractive interfaces, where $N$ is the number of
dielectric elements. The ray density is chosen such that neighboring
points on each surface are separated by much less than an optical
wavelength.

We start by designing the first dielectric surface that light meets
after propagating in vacuum. More specifically, we start from the point
where the first interface meets the 0° ray, on the optical axis. As this
ray is already collimated along the optical axis, it needs no
correction, setting the interface perpendicular to the optical axis. We
then move to the ray at a small angle $\varepsilon$ from the axis, which must be
tilted by $\varepsilon$ divided by the number of interfaces. This is done by using
Snell's law to find the surface slope that gives this tilt and extending
the surface along that slope until it meets the next ray (fig.~\ref{fig:S1}A). We
then repeat this process, ray by ray, until reaching the 180° ray (fig.~\ref{fig:S1}B).
The rays leaving the first interface are the input for the second
interface, which we design in the same way, and so on (fig.~\ref{fig:S1}C). The
routine is fully parametric -- refractive index, number of shells, ray
density, and weighting parameters are free inputs. Fig.~\ref{fig:S2} summarizes
the full workflow as a flowchart. The source code and compiled version
of the routine are available at
\url{https://github.com/na2radiant-hash/NA2-lens-designer} (release
v1.1); running the script LensConstructionGradual.m regenerates every
lens and simulated figure panel of the paper.

\subsection{Collection efficiency of a dipole emitter}
\label{sec:dipole}

To quantify the cost of the access opening (Fig.~\ref{fig:network}, B and C), we
consider a dipole aligned with the optical axis, whose radiated
intensity per unit solid angle is $I(\alpha) \propto \sin^{2}\alpha$,
where $\alpha$ is the polar angle measured from the optical axis, as for the
launch angles $\alpha_{0}$ above. Light emitted into a cone of half-angle $\theta$ about
the backward axis ($\alpha = 180^{\circ}$) escapes through the opening uncollected,
and all other light is collected. The full apex angle of this cone is
$2\theta$, and its numerical aperture is $\mathrm{NA} = \sin\theta$. The collection efficiency
$\eta$ is therefore
\begin{equation}
  \eta(\theta) = 1 - \frac{\int_{\pi - \theta}^{\pi} \sin^{3}\alpha \, \mathrm{d}\alpha}{\int_{0}^{\pi} \sin^{3}\alpha \, \mathrm{d}\alpha}
  \label{eq:eta-integral}
\end{equation}
Using
$\int_{0}^{\theta} \sin^{3}\alpha \, \mathrm{d}\alpha = \frac{2}{3} - \cos\theta + \frac{\cos^{3}\theta}{3}$
and $\int_{0}^{\pi} \sin^{3}\alpha \, \mathrm{d}\alpha = \frac{4}{3}$, we obtain
\begin{equation}
  \eta(\theta) = \frac{2 + 3\cos\theta - \cos^{3}\theta}{4}
  \label{eq:eta-theta}
\end{equation}
or, in terms of the numerical aperture of the opening,
\begin{equation}
  \eta(\mathrm{NA}) = \frac{2 + 3\sqrt{1 - \mathrm{NA}^{2}} - \left( 1 - \mathrm{NA}^{2} \right)^{3/2}}{4}
  \label{eq:eta-na}
\end{equation}
For a small opening, Eq.~\ref{eq:eta-na} reduces to
\begin{equation}
  1 - \eta \approx \frac{3}{16}\,\mathrm{NA}^{4}
  \label{eq:small-na}
\end{equation}
The loss is quartic rather than quadratic in NA because the dipole does
not radiate along its axis; an isotropic emitter would lose $\mathrm{NA}^{2}/4$
through the same opening. Fig.~\ref{fig:network}C plots Eq.~\ref{eq:eta-theta} as a function of the apex
angle $2\theta$, evaluated analytically rather than by ray tracing. Collection
efficiencies of 99.9\% and 99\% are reached at $\theta = 15.6^{\circ}$ ($2\theta = 31.2^{\circ}$, $\mathrm{NA}
= 0.27$) and $\theta = 28.1^{\circ}$ ($2\theta = 56.2^{\circ}$, $\mathrm{NA} = 0.47$), respectively.

\subsection{Fabricability of a three-dimensional Allight collector}
\label{sec:fabricability}

Allight requires transparent materials. The material must also combine a
refractive index high enough for a compact collector (near 1.7) with
high transmission over a broad spectrum. Sapphire is one example that
meets these requirements: its refractive index is near 1.77, and it
transmits the key wavelengths of quantum optics with trapped ions and
with neutral atoms such as ytterbium and erbium, for excitation,
collection, optical tweezers, and cooling (from 280~nm for the magnesium
ion to 1.76~µm for the barium ion), and an even wider range, from 0.17
to 5~µm, for future applications. Because of the internal void, we
suggest making the first element from two parts, then optically aligning
and bonding them. Developing Allight faces challenges, including
polishing aspheric surfaces in materials such as sapphire and aligning
the two parts of the first element around the internal void; current
technology addresses both ($\lambda/2$ catalog aspheres; $\lambda/4$ off-the-shelf
alignment). Each of the eight sapphire interfaces also needs an
antireflection coating; assuming a coating that transmits 99.99\% per
interface, the total transmission loss is less than 0.1\% ($0.9999^{8} =
0.9992$). Companies such as Edmund Optics and Coherent list such aspheres
in their catalogs, and Opto-Alignment Technology, Inc.\ is among the
companies that provide such alignment, as listed in their catalogs.

To test these tolerances, we perturbed a sapphire version of the
collector ($n = 1.77$, four elements, designed with the algorithm above)
in two ways: we added to each of its eight interfaces a random smooth
height error of $\lambda/2$ peak-to-valley ($\lambda = 632.8$~nm), applied along the
surface normal with a correlation length $l_{\mathrm{c}}$ of 0.5, 1,
2, or 5~mm, and, separately, we offset the rear part of the first
element by $\lambda/4$ transversely and axially, counting rays that reach the
resulting step as lost. We then traced rays of the axial dipole of Fig.~\ref{fig:network}
through the perturbed collector and counted the power that falls
outside the design beam envelope, over 100 random realizations for each
$l_{\mathrm{c}}$. The figure error deflects the rays by up to 0.2~mrad
rms ($l_{\mathrm{c}} = 0.5$~mm), which blurs the beam edge by up to
0.2~mm at 1~m; the resulting loss at 1~m is at most 0.006\% on average
and 0.03\% in the worst realization. The $\lambda/4$ offset loses 0.004\% at any
distance, and even a full-wave offset loses only 0.016\%. The loss is
small because the blur is only 0.2~mm on a 200-mm beam and because the
beam edge carries the weak backward emission of the dipole (fig.~\ref{fig:S3}).
Both losses are therefore well below 0.1\%. Hence, the extension of the
current state of the art that Allight needs is comparable to the
extensions that introduced other unconventional optics: freeform optics,
which transformed microscope objectives to broadband operation
\cite{Bauer2018,Rolland2021} and lithography to the extreme ultraviolet, as
well as metamaterial lenses \cite{Bomzon2002}, which transformed lidars (light
detection and ranging) from large devices to cell phones.

\section{Acknowledgments}
\label{sec:acknowledgments}

We thank R. Shmuel and D. Bitolin (School of Mechanical Engineering, Tel Aviv
University) for fabricating the polycarbonate collector shown in
Fig.~\ref{fig:experiment}. We also thank N. Aton for assistance with the graphic
design of Fig.~\ref{fig:design}.

\medskip
\noindent\textbf{Funding:} This work was supported by the United States--Israel
Binational Science Foundation (NSF-BSF) (2020683) and the Israel Science Foundation
(537/20 and 250/26) and by Israel Science Foundation research grant (ISF’s
No.~4098/25) and the Maimonides Fund’s Future Scientists Center.

\clearpage
\section{Supplementary Figures}
\label{sec:supplementary}

\setcounter{figure}{0}
\renewcommand{\thefigure}{S\arabic{figure}}
\renewcommand{\theHfigure}{S\arabic{figure}}

\begin{figure}[!ht]
  \centering
  \includegraphics[width=0.85\linewidth]{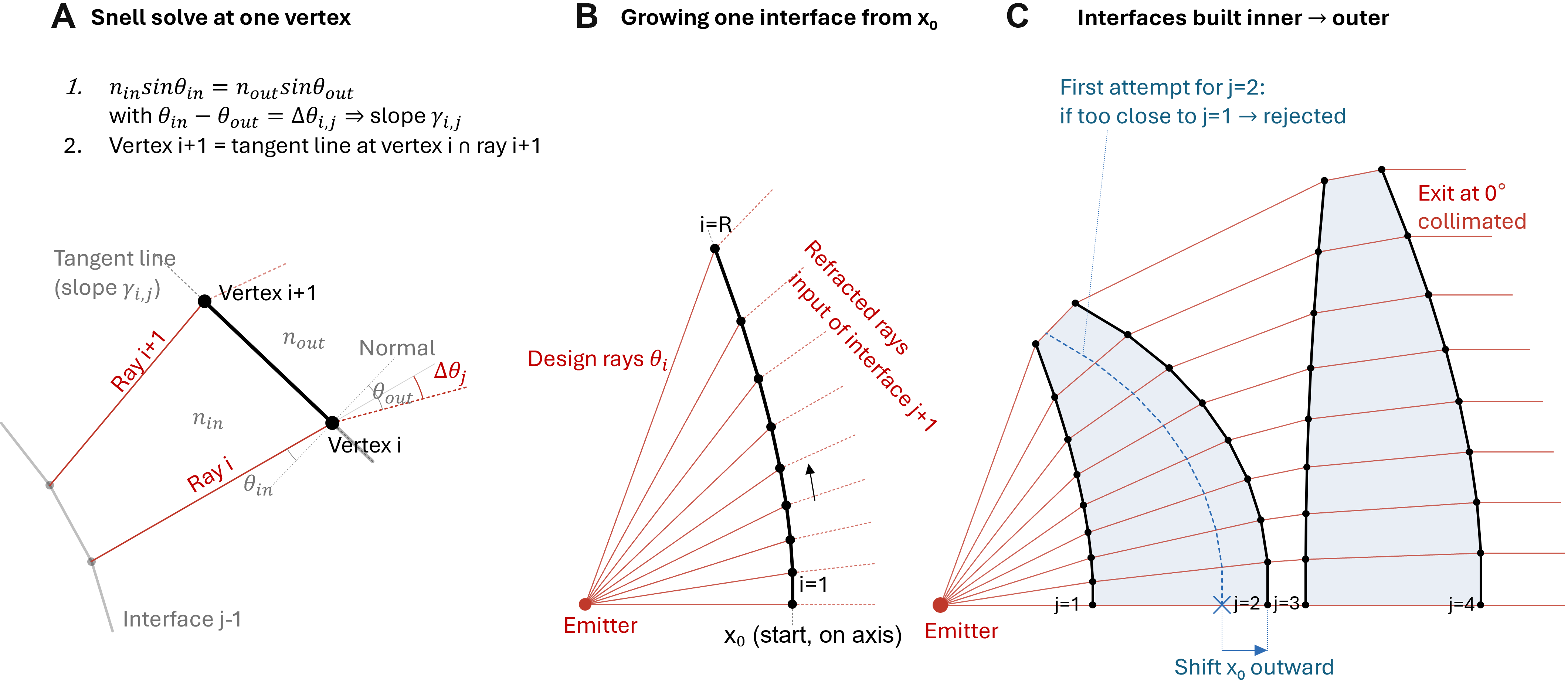}
  \caption{\textbf{Inverse-design algorithm.} (A) Construction of one
  surface point. At vertex $i$ of interface $j$, Snell's law sets the surface
  slope that tilts ray $i$ by its assigned deflection $\Delta\theta(i, j)$; the next
  vertex, $i + 1$, lies where the tangent line through vertex $i$ meets ray
  $i + 1$. (B) Growing one interface. Starting on the optical axis at $x_{0}$,
  where the surface is perpendicular to the axis, vertices are added ray
  by ray from $i = 1$ to the 180° ray, $i = R$. The refracted rays are the
  input of the next interface. (C) Building the collector from the inside
  out. Interfaces $j = 1$ to 4 bound the glass elements (shaded), and all
  rays exit collimated at 0°. An interface that comes too close to the
  previous one, such as the first attempt for $j = 2$ (blue dashed line), is
  rejected, and its starting point $x_{0}$ is shifted outward.}
  \label{fig:S1}
\end{figure}
\clearpage

\begin{figure}[!t]
  \centering
  \includegraphics[width=0.6\linewidth]{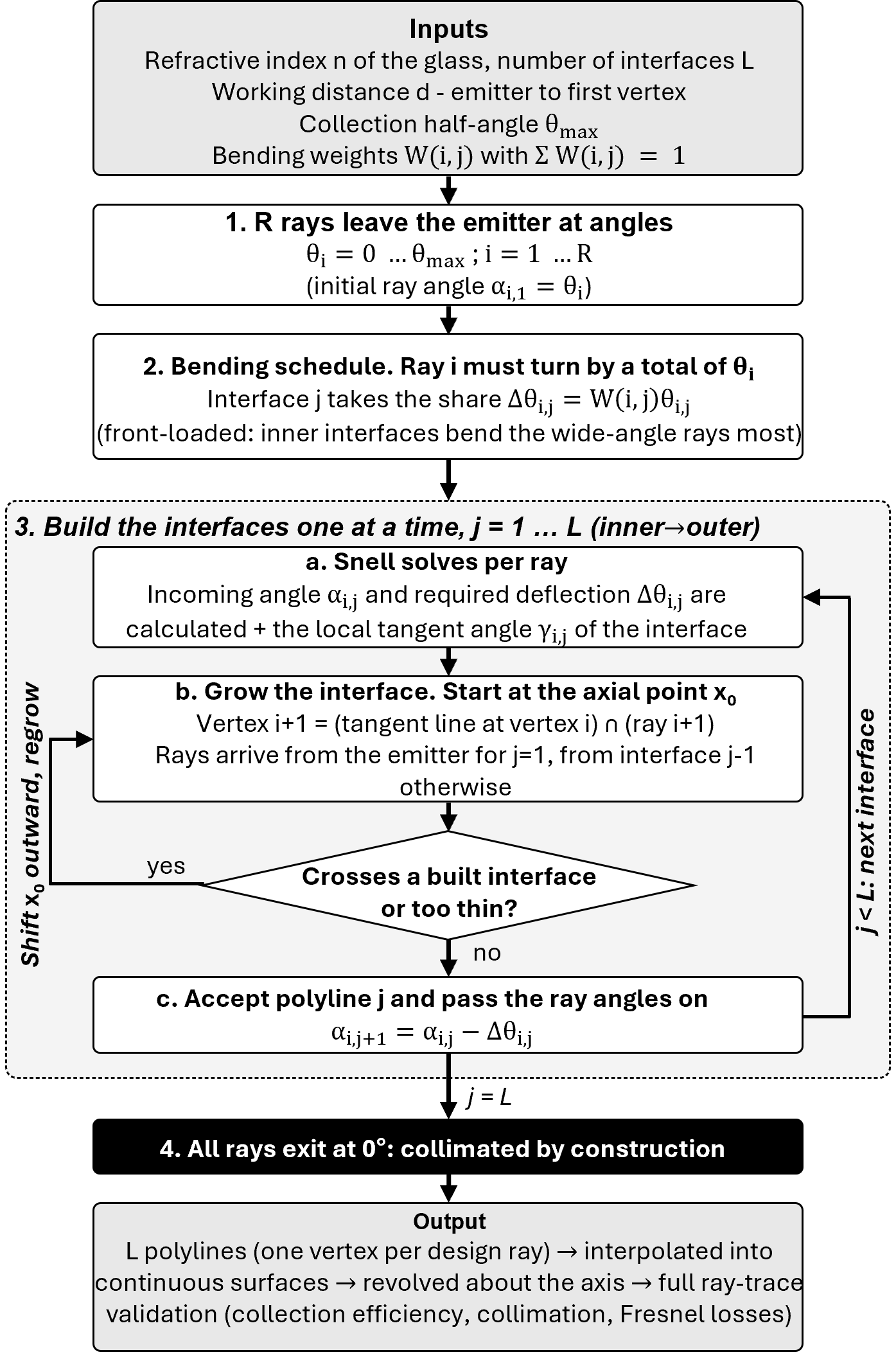}
  \caption{\textbf{Flowchart of the inverse-design algorithm.} Inputs are
  the refractive index $n$, the number of interfaces $L$, the working distance
  $d$, the collection half-angle $\theta_{\mathrm{max}}$, and the bending
  weights $W(i, j)$, with $\Sigma_{j} W(i, j) = 1$. (1) A fan of $R$
  design rays leaves the emitter at angles $\theta_{i}$ from 0 to
  $\theta_{\mathrm{max}}$. (2) Each ray $i$ must turn by $\theta_{i}$ in
  total; interface $j$ takes the share $\Delta\theta_{i,j} = W(i, j)\,
  \theta_{i}$. (3) The interfaces are built one at a time, from the
  inside out ($j = 1$ to $L$): (a) Snell's law gives the local tangent angle
  $\gamma_{i,j}$ that produces the required deflection; (b) the
  interface is grown vertex by vertex from the axial point $x_{0}$; if it
  crosses an existing interface or is too thin, $x_{0}$ is shifted outward and
  the interface is regrown; (c) otherwise, it is accepted and the ray
  angles $\alpha_{i,j+1} = \alpha_{i,j} - \Delta\theta_{i,j}$ are passed to the next interface. (4) After the
  last interface ($j = L$), all rays exit at 0°, collimated by construction.
  The resulting $L$ polylines are interpolated into continuous surfaces,
  revolved about the optical axis, and validated by full ray tracing of
  collection efficiency, collimation, and Fresnel losses.}
  \label{fig:S2}
\end{figure}
\clearpage

\begin{figure}[!t]
  \centering
  \resizebox{0.85\linewidth}{!}{%
    \setlength{\unitlength}{1cm}%
    \begin{picture}(15.907,13.760)
      \put(0.085,0.302){\includegraphics[width=8.513cm]{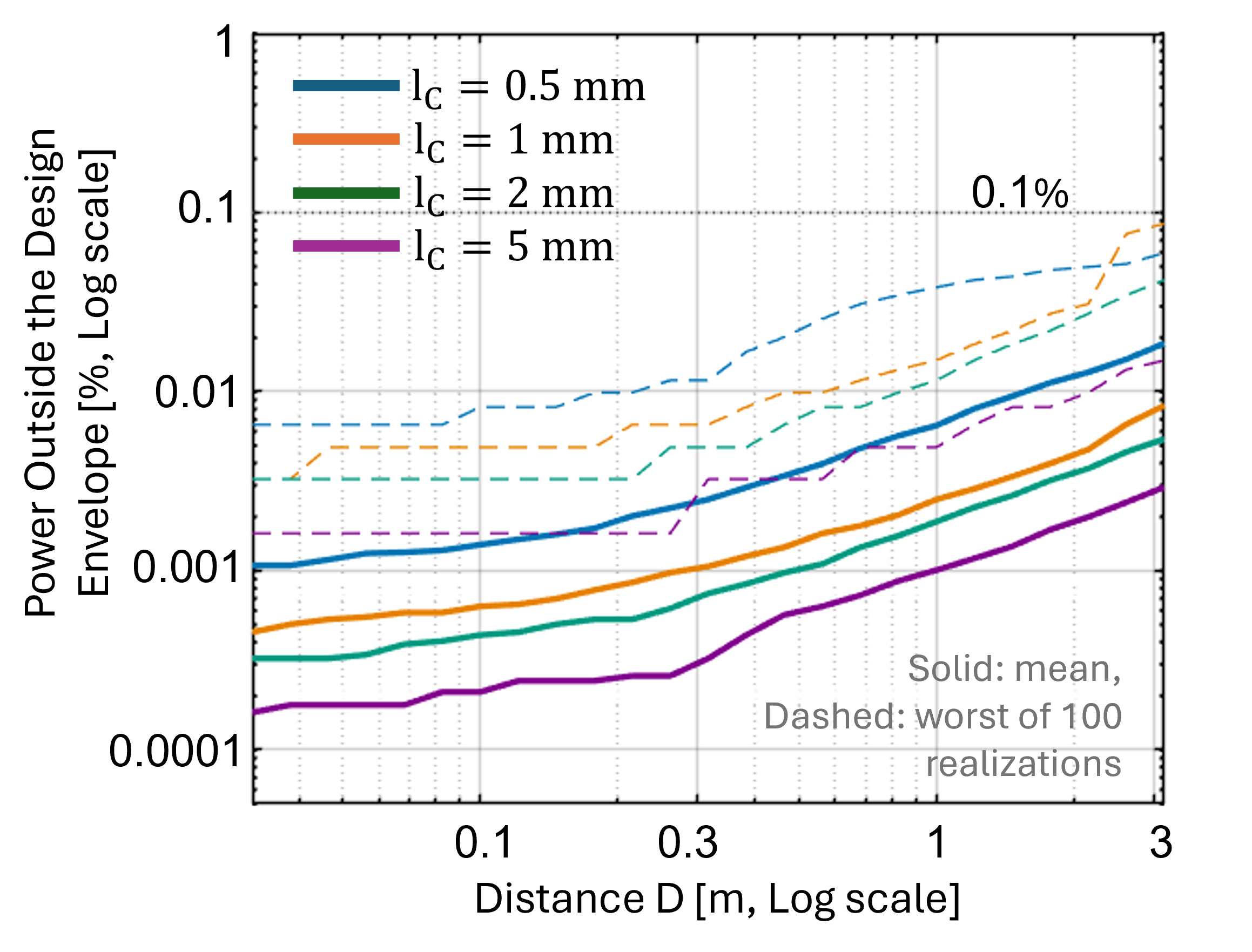}}
      \put(2.963,6.856){\includegraphics[width=9.408cm]{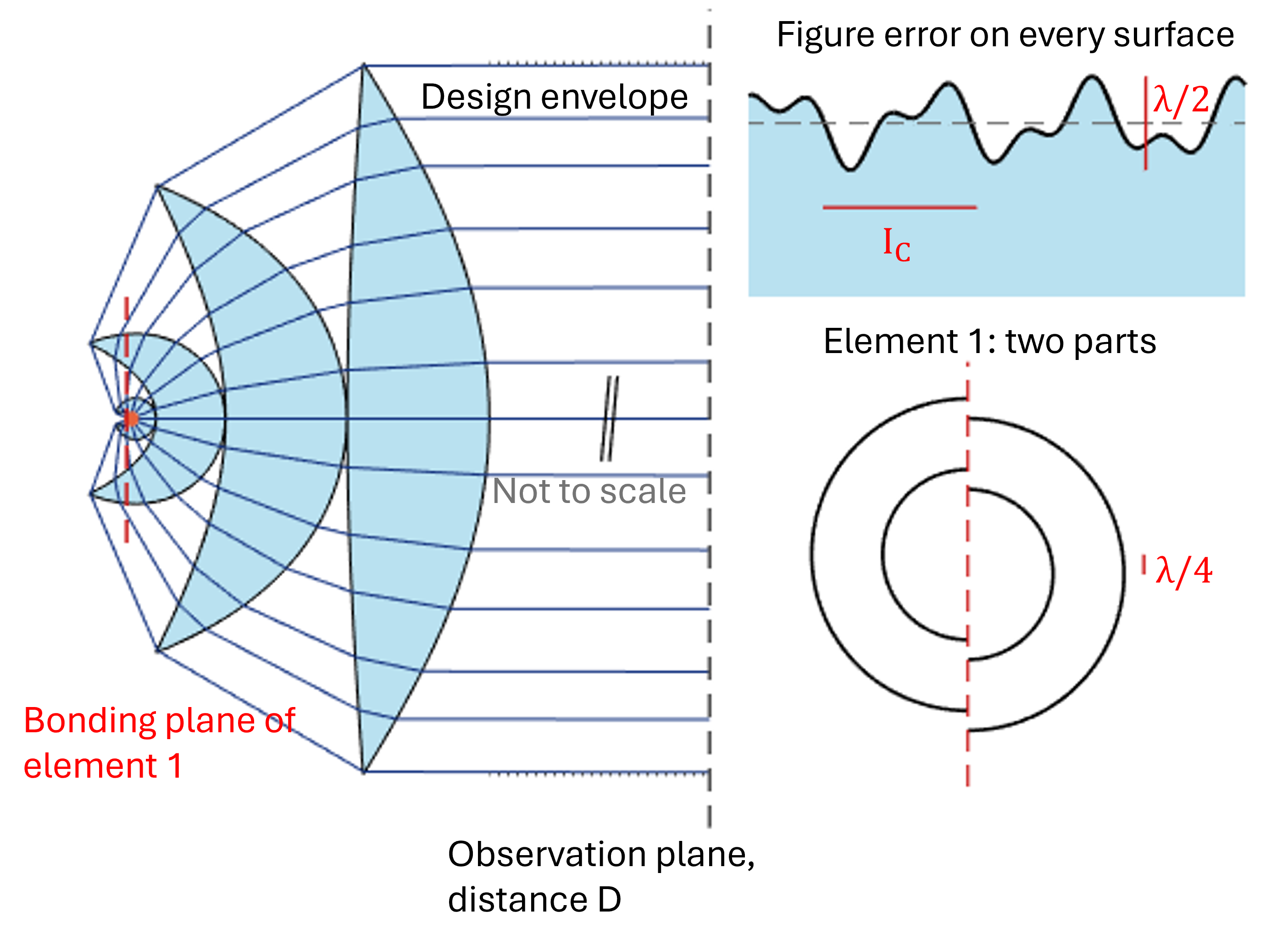}}
      \put(8.042,0.000){\includegraphics[width=7.865cm]{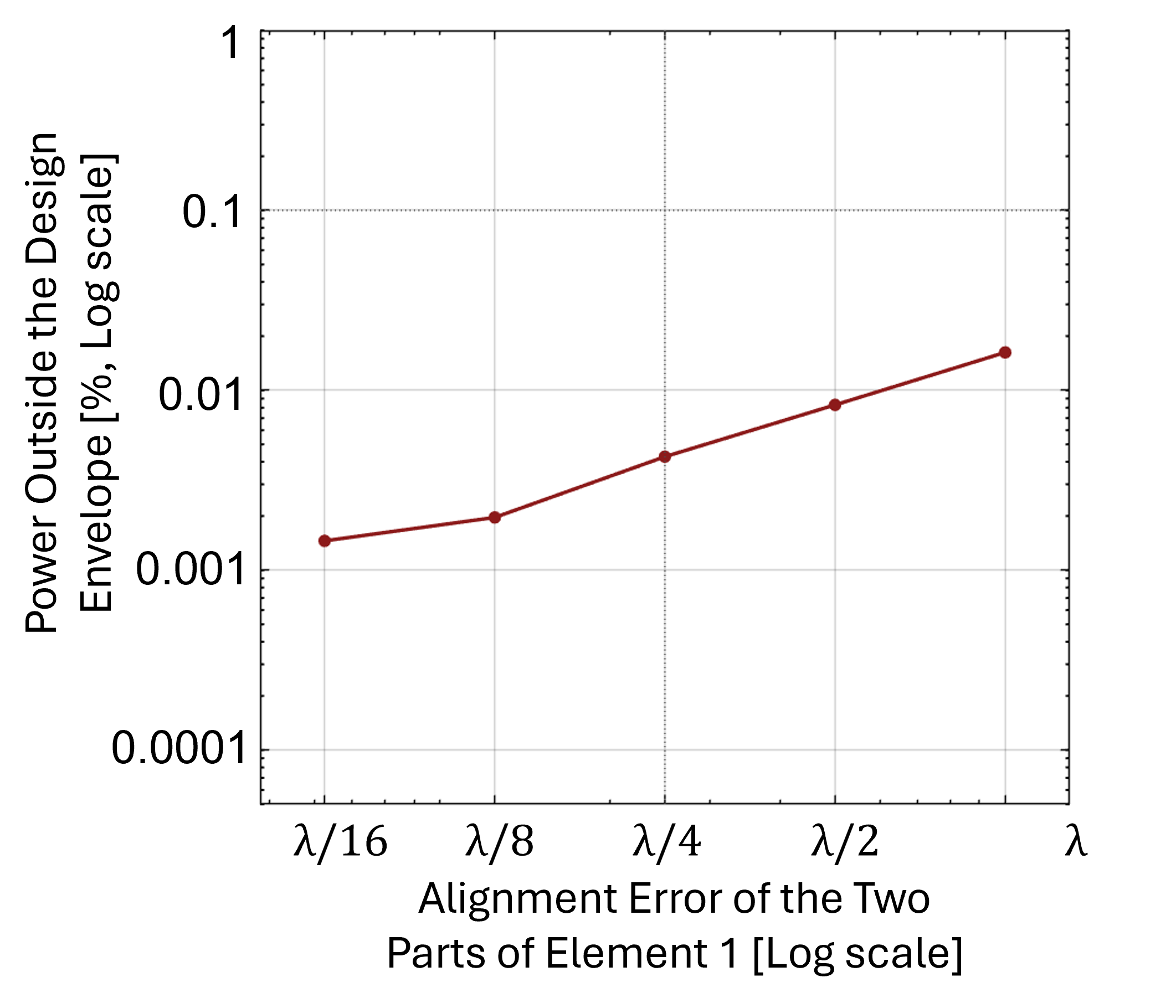}}
      \put(0,0){\includegraphics[width=15.907cm]{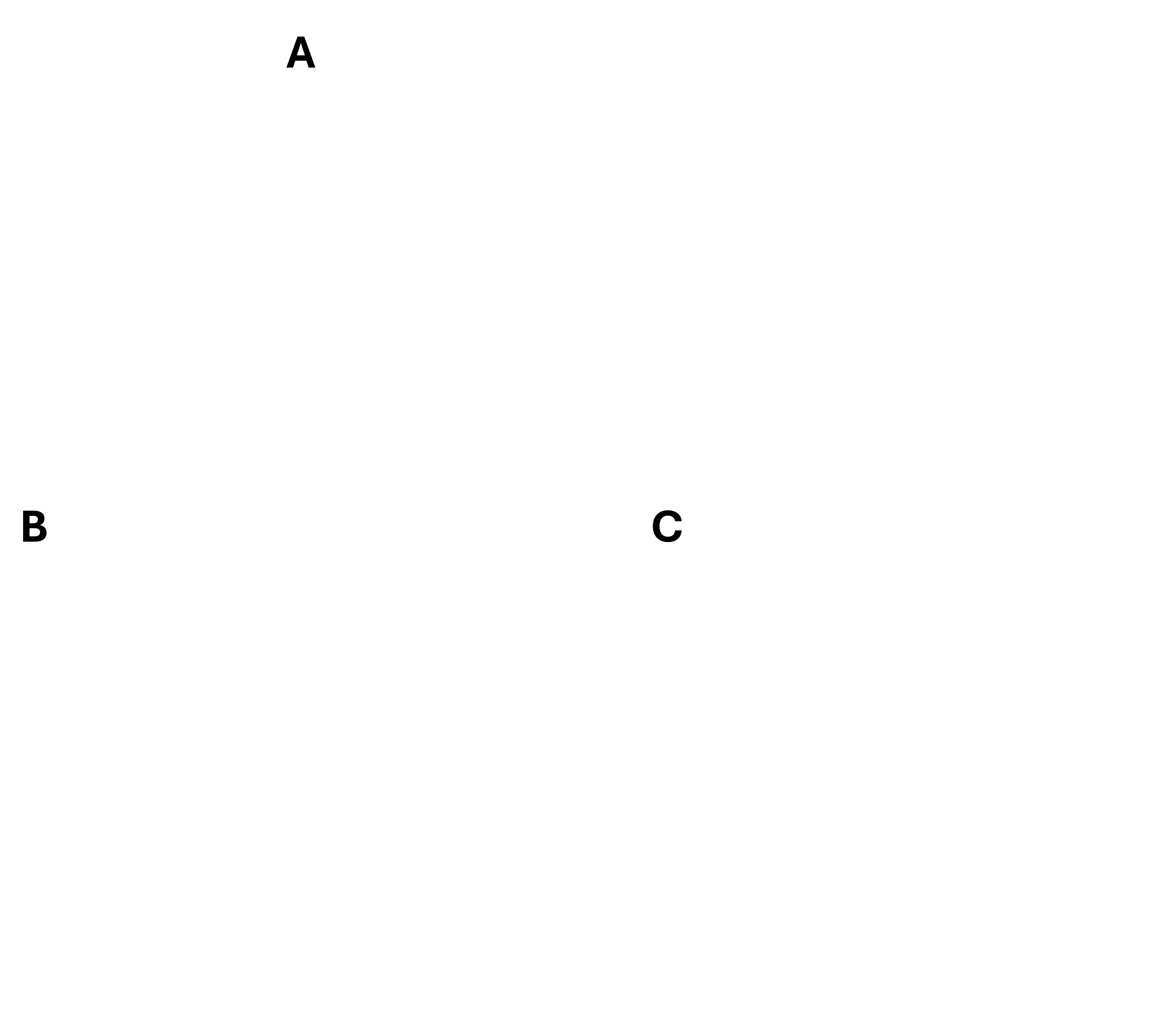}}
    \end{picture}}
  \caption{\textbf{Fabrication-tolerance check.} (A) The design collector
  with its rays, the bonding plane of the first element (red), and the
  observation plane at distance $D$ with the design beam envelope (not to
  scale). Insets show the two perturbations, exaggerated: a $\lambda/2$
  peak-to-valley figure error with correlation length $l_{\mathrm{c}}$
  on each of the eight interfaces, and the rear part of the first element
  offset by $\lambda/4$ transversely and axially. (B) Power outside the design
  envelope versus $D$ for the figure error with $l_{\mathrm{c}} = 0.5$, 1,
  2, and 5~mm, for the axial dipole of Fig.~\ref{fig:network}; solid lines, mean; dashed
  lines, worst of 100 realizations at each $D$; dotted line, 0.1\%. (C)
  Power outside the design envelope versus the alignment error of the two
  parts of the first element, applied as an offset of that size
  transversely and axially; solid line, mean of the two transverse
  directions; values at $D = 1$~m.}
  \label{fig:S3}
\end{figure}

\end{document}